**Controllable spin-polarization using external electric field and spin-inversion in a silicene quantum dot**

Hamidreza Simchi[*,1,2], Mahdi Esmaeilzadeh[**,1], Hossein Mazidabadi[1] and Mohammad Norouzi[1]

[1]Department of Physics, Iran University of Science and Technology, Narmak, Tehran 16844, Iran

[2]Semiconductor Technology Center, Tehran Iran

We study spin dependent electron transport properties in a silicene quantum dot using tight-binding non-equilibrium Green's function method. It is shown that, for appropriate values of electron energy, the silicene dot can work as a controllable spin polarizer. The spin polarizer can polarize the spin of transmitted electrons from nearly pure down to nearly pure up by changing the strength of an external electric field. Also, for spin polarized incoming electrons, the silicene quantum dot can invert the spin of electrons and can works as a spin inverter or a spin NOT gate. In addition, we investigate the effects of exchange field, induced by ferromagnetic substrate, on electron conductance and show that the silicene quantum dot can act as a nearly perfect spin-filter in the presence of exchange field.

## I. Introduction

Field effect transistor (FET) is a key element in microelectronics. In a FET the (charge) current between its source and drain can be controlled by applying a gate voltage.[1] A single molecule n-type FET-like device has been made by using perylene tetracarboxylic dimide.[2] Scientists search new materials and methods for adding the spin degrees of freedom to conventional (charge-based) electronic devices. Spin injection, transport, control and manipulation, as well as detection of spin polarization play an important role in spin-based electronics (or spintronics).[3] Spin polarized field effect transistor (spin FET) has been proposed by Datta and Das in 1990.[4] In the spin FET, the spin polarized electrons are injected from the source into a non-magnetic region and then it transmitted to the drain.[4] Recently, spin dependent electron transport properties such as spin filtering, spin polarization, and spin inversion have been studied vastly (see e.g., Refs. 5-12).

Silicene, the graphene counterpart of silicon[13], has a buckled structure with lattice constant of 3.8Å and buckling distance of 0.406Å.[14] The low energy effective Hamiltonian involving spin-orbit coupling in silicene has been introduced by Liu et al.[15] It has been shown that, although the nearest-neighbor spin-orbit interaction (SOI) vanishes, the next-nearest-neighbor SOI does not vanish due to the buckled nature of silicene lattice.[15, 16] In consequence, the Hamiltonian of silicene includes not only the intrinsic SOI term but also the intrinsic Rashba SOI term.[15, 16] The topological insulating properties[17], intrinsic spin Hall effect[18], induced spin polarized current,[19, 20] field effect topological quantum transistor[21], photo-induced topological phase transition,[22] and graphene-silicene-graphene heterojunction[23] have been studied.

In this paper, spin-dependent transport properties in a silicene quantum dot are studied using tight-binding non-equilibrium Green's function method. To manipulate and control the spin polarization of transmitted electrons, an external electric field is applied perpendicularly to the plane of the dot. It is shown that the silicene dot can work as a controllable spin polarizer. Furthermore, the silicene quantum dot can work as a spin inverter or spin NOT gate. The organization of paper is as follow. In section II, the calculation method and in section III, the results and discussion are presented. Finally a summary and conclusion are given in section IV.

## II. Calculation method

We use tight binding non-equilibrium Green's function (TB-NEGF) method to study the spin dependent electron transport properties of buckled zigzag silicene quantum dot connected to two semi-infinite silicene nanoribbon as left and right leads (see Fig. 1). The Hamiltonian of dot can be written as[23, 24,16]

$$H_C = -t \sum_{<ij>,\alpha} c_{i\alpha}^\dagger c_{j\alpha} - i\frac{2\lambda_R}{3} \sum_{<<i,j>>,\alpha\beta} \mu_i c_{i\alpha}^\dagger (\sigma \times \hat{d}_{ij})_{\alpha\beta}^z c_{j\beta} + i\frac{\lambda_{SO}}{3\sqrt{3}} \sum_{<<ij>>,\alpha\beta} v_{ij} c_{i\alpha}^\dagger \sigma_{\alpha\beta}^z c_{j\beta}$$
$$+ l_z E_z \sum_{i,\alpha} \xi_i c_{i\alpha}^\dagger c_{i\alpha} + M \sum_{i\alpha} c_{i\alpha}^\dagger \sigma_z c_{i\alpha} + h.c.$$
(1)

where $t$ is the nearest neighbor hopping integral, $\prec ij \succ$ and $\prec\prec ij \succ\succ$ denote all the nearest neighbor (NN) and next nearest neighbor (NNN) hopping sites, respectively, $\alpha$ and $\beta$ label the spin quantum numbers, and the operator $c_{i\alpha}^\dagger$ ($c_{j\alpha}$) creates (annihilates) an electron with spin $\alpha$ at site $i$ ($j$). In Eq. (1), the second term describes the Rashba SOI between NNN sites where $\mu_i = \pm 1$ for the A (B) site, $\sigma = (\sigma_x, \sigma_y, \sigma_z)$ is the vector of the Pauli matrix, $\hat{d}_{ij} = \vec{d}_{ij}/|\vec{d}_{ij}|$ is the unit vector of $\vec{d}_{ij}$ which connects NNN sites $i$ and $j$. The third term represents the intrinsic SOI between NNN sites where $v_{ij} = -1(+1)$ if the NNN hopping is clockwise (anticlockwise) with respect to the positive z axis, $2l_z = 0.46$ Å is buckling distance[23] and $E_z$ (in the forth term) is the strength of an external electric field in eV/Å. In the fifth term, M is the exchange field induced by ferromagnetic substrates. Equation (1) can be written as[17,19]

$$H_C \propto \begin{pmatrix} -\lambda_{SO} + lE_z + M & -t & ia\lambda_R k_- & 0 \\ -t & \lambda_{SO} - lE_z + M & 0 & -ia\lambda_R k_- \\ -ia\lambda_R k_+ & 0 & \lambda_{SO} + lE_z - M & -t \\ 0 & ia\lambda_R k_+ & -t & -\lambda_{SO} - lE_z - M \end{pmatrix} \qquad (2)$$

in the basis $\{\psi_A^\uparrow, \psi_B^\uparrow, \psi_A^\downarrow, \psi_B^\downarrow\}^T$, where $\uparrow(\downarrow)$ means spin up (down), $T$ means transpose, $k_\pm = k_x \pm k_y$, and $a = 3.86$ Å.[17] It should be noted that we ignore the extrinsic Rashba SOI which is induced by external electric field because of its small value (i.e., of order $10^{-6}$ eV).[22] Since, the band gap of silicene closes at the critical electric field $|E| = E_{cr} = \frac{\lambda_{SO}}{l} = 0.017$ eV/Å,[17,22], the absolute value of applied electric field is always greater than the value of critical field in our numerical calculations. Also, for simplicity, we assume that the intrinsic and intrinsic Rashba SOI are zero in the leads.[14,17] In other words, we assume that the structure of leads is planar which leads to zero SOI coefficients.[14,17] In the present study, we choose t=1.6 eV, $\lambda_R = 0.5t$, and $\lambda_{SO} = 0.5t$.[23] The Hamiltonian of leads is given by[25]

$$H_{L(R)} = -t \sum_{\langle ij \rangle} d_i^\dagger d_j, \qquad (3)$$

where the operator $d_i^\dagger$ ($d_j$) creates (annihilates) an electron in leads. The spin transport properties of the elongated silicone device (i.e., quantum dot) is studied using a divide and conquer tight binding approach.[26] The interaction matrices between the ribbon and left and right leads can be written as[25,26]

$$H_{CL} = \begin{pmatrix} 0 & -t & 0 & 0 & \cdots & \cdots & 0 \\ 0 & 0 & 0 & 0 & \cdots & \cdots & 0 \\ 0 & 0 & 0 & 0 & \cdots & \cdots & 0 \\ 0 & 0 & -t & 0 & \cdots & \cdots & 0 \\ 0 & 0 & 0 & 0 & 0 & 0 & 0 \\ \vdots & \vdots & \vdots & \vdots & \vdots & \vdots & \vdots \\ \vdots & \vdots & \vdots & \vdots & \vdots & \vdots & \vdots \end{pmatrix}, \quad H_{CR} = \begin{pmatrix} 0 & 0 & 0 & 0 & \cdots & \cdots & 0 \\ -t & 0 & 0 & 0 & \cdots & \cdots & 0 \\ 0 & 0 & 0 & -t & 0 & \cdots & 0 \\ 0 & 0 & 0 & 0 & \cdots & \cdots & 0 \\ 0 & 0 & 0 & 0 & 0 & 0 & 0 \\ \vdots & \vdots & \vdots & \vdots & \vdots & \vdots & \vdots \\ \vdots & \vdots & \vdots & \vdots & \vdots & \vdots & \vdots \end{pmatrix}, \quad (4)$$

respectively. In fact, we consider a linear chain of unit cells such that the unit cells of left lead (L) are placed at positions $-\infty$.....-3, -2, -1 the ribbon (C) is placed at positions 0, 1, 2, .....,n-1,n, and the unit cells of right lead (R) are placed at positions n+1, n+2,n+3,....... $+\infty$. The surface Green's function of left ($g_L$) and right ($g_R$) leads are calculated by suing the Sancho's algorithm.[27] The self energies and the coupling matrices can be calculated by using[22]

$$\sum_L = H_{CL} g_L H_{CL}^\dagger, \qquad \sum_R = H_{CR} g_R H_{CR}^\dagger \qquad (5)$$

and

$$\Gamma_{L(R)} = i\left(\sum_{L(R)} - \sum_{L(R)}^\dagger\right) \qquad (6)$$

where $\sum_{L(R)}$ and $\Gamma_{L(R)}$ are self energy and coupling matrices of left (right) lead, respectively. The Green's function of the ribbon can be calculated by using[25]

$$\xi = \begin{pmatrix} \xi^{\uparrow\uparrow} & \xi^{\uparrow\downarrow} \\ \xi^{\downarrow\uparrow} & \xi^{\downarrow\downarrow} \end{pmatrix} = ((E + i\eta) \times I - H_C - \sum_{NL} - \sum_{NR})^{-1} \qquad (7)$$

where $E$ is electron energy, $\eta$ is an infinitesimal number, $I$ is the unit matrix, and $\sum_{NL(NR)}$ is defined by:[25,28,29]

$$\sum_{NL(NR)} = \begin{pmatrix} \sum_{L(R)} & 0 \\ 0 & \sum_{L(R)} \end{pmatrix}. \tag{8}$$

Now, the spin dependent conductance can be calculated by using[25,28,29]

$$\begin{aligned} G^{\uparrow\uparrow} &= \mathrm{Im}\,(Trace(\Gamma_L \times \xi^{\uparrow\uparrow} \times \Gamma_R \times \xi^{\uparrow\uparrow\dagger})), \\ G^{\uparrow\downarrow} &= \mathrm{Im}\,(Trace(\Gamma_L \times \xi^{\uparrow\downarrow} \times \Gamma_R \times \xi^{\uparrow\downarrow\dagger})), \\ G^{\downarrow\uparrow} &= \mathrm{Im}\,(Trace(\Gamma_L \times \xi^{\downarrow\uparrow} \times \Gamma_R \times \xi^{\downarrow\uparrow\dagger})), \\ G^{\downarrow\downarrow} &= \mathrm{Im}\,(Trace(\Gamma_L \times \xi^{\downarrow\downarrow} \times \Gamma_R \times \xi^{\downarrow\downarrow\dagger})), \end{aligned} \tag{9}$$

where $G^{\sigma\sigma'}$ is the electron conductance with incoming spin $\sigma$ and outgoing spin $\sigma'$, $\sigma, \sigma' = \uparrow, \downarrow$, and Im denotes the imaginary part. The spin-polarization is defined by[25,28,29]

$$P_S \equiv \frac{(G^{\uparrow\uparrow} + G^{\downarrow\uparrow}) - (G^{\downarrow\downarrow} + G^{\uparrow\downarrow})}{(G^{\uparrow\uparrow} + G^{\downarrow\uparrow}) + (G^{\downarrow\downarrow} + G^{\uparrow\downarrow})} \tag{10}$$

Another quantity which is useful to study spin flip (or spin rotation) probability of the transmitted electrons can be defined by

$$P_{sf} \equiv \frac{G_{sf}}{G_{sc} + G_{sf}} \tag{11}$$

where $G_{sf} \equiv G^{\uparrow\downarrow}+G^{\downarrow\uparrow}$ is the spin-flipped conductance of transmitted electrons and $G_{sc}=G^{\uparrow\uparrow}+G^{\downarrow\downarrow}$ is the spin-conserved conductance.[30] The total conductance is the sum of spin-conserved conductance and spin-flipped conductance i.e., $G=G_{sc}+G_{sf}$.

## III. Results and discussion

In this section, numerical results of spin dependent electron transport properties in a silicene ribbon are presented and discussed. The ribbon composed of ten unit cells (with total 120 atoms) as shown in Fig. 1.

Figures 2(a) and 2(b) show the spin-dependent conductance of transmitted electrons versus the electron energy. It is observed that the conductance of electrons with spin flip from down to up i.e., $G^{\downarrow\uparrow}$ and inversely the conductance of electrons with spin flip from up to down i.e., $G^{\uparrow\downarrow}$ are not zero [see Figs. 2(a) and 2(b), respectively] except for some special values of electron energy. This is an important effect which is due to the intrinsic SOI and intrinsic Rashba SOI of silicene ribbon. In Fig. 2(c), the spin-polarization of transmitted electrons $P_s$ versus the electron energy is shown. As Fig. 2(c) shows, spin polarization changes with changing the electron energy. For some electron energy regions, spin polarization $P_s$ is negative which shows that unpolarized electrons can transmit through the silicene ribbon with down spin polarization. Also, spin polarization $P_s$ is positive for some other energy regions which shows that unpolarized electrons can transmit with up spin polarization. Note that spin polarization changes from -0.63 to +0.58 and is not perfect because it cannot reach to -1 and +1 [see Fig. 2(c)].

To manipulate and control the spin polarization of transmitted electrons, we now consider an external electric field applied perpendicularly to the plane of the ribbon. In Fig. 3(a), the spin polarization $P_s$ versus the electric field is shown for specific values of electron energy. As we can see in this figure, for different values of electron energy, the spin polarization can be changed by changing the electric field. The changing of spin polarization by electric field for electron energy $E - E_F = 0$ is small (it changes from 0.16 to 0.24), but for $E - E_F = 0.2eV$ is relatively large (it changes from -0.33 to 0.80). To obtain an optimum value of electron energy, a three-dimensional graph of spin polarization versus the electric field and electron energy is shown in Fig. 3(b). It is seen that there are many electron energy values at which the variation of spin polarization is large e.g., near the electron energy value $E - E_F = +0.72$, the spin polarization changes from $P_s = -0.65$ to $P_s = +0.45$ by changing the electric field. Therefore, the silicene ribbon can work as a controllable (or tunable) spin polarizer. It should be noted that one can change the electric field by using a gate voltage.

Figure 4 shows a three-dimensional graph of spin flipping probability $P_{sf}$ versus the electron energy and electric field. For definition of spin flipping probability, see Eq. (11). As shown in this figure, there are some values of electron energy and electric field at which spin flipping with high probability ($P_{sf} > 0.9$) can take place. This is an interesting result because if one uses incoming electrons with a specific spin polarization, the silicene ribbon can invert the spin of transmitted electrons from up to down and vice versa. Therefore, the silicene ribbon can also work as a spin inverter or a spin NOT gate.

The effect of exchange field M on spin polarization is shown in Figs. 5(a) and 5(b) for electron energy values $E - E_F = -0.7$ and $+0.2eV$, respectively. The exchange field can be generated by depositing of silicene on a ferromagnetic substrate.[17,19,31] As these figures show, the spin polarization can also change due to the exchange field. An interesting result is related to curve of $M=0.64$ and $E-E_F=-0.7\text{eV}$ [see Fig. 5(a)] which shows that the spin polarization is very close to +1 independent of electric field. It means that, for this case, the silicene ribbon can work as a nearly perfect spin filter. It should be noted that, different values of $\lambda_R$, and $\lambda_{SO}$ have been reported. [16,23,32] For, $\lambda_R=8.66\,\text{meV}$, and $\lambda_{SO}=4.2\,\text{meV}$[32] the Fig.6 shows the spin polarization versus electron energy and electric field. As it shows near Fermi energy spin inverting can be happened. Fig.7 shows the conductance versus electron energy and electric field for $\lambda_R=8.66\,\text{meV}$, and $\lambda_{SO}=4.2\,\text{meV}$[32] By attention to Fig.6 and Fig.7, it can be concluded that the spin inverting can be happened only near Fermi energy since for other electron energy values the $G^{\uparrow\uparrow}$ and $G^{\downarrow\downarrow}$ are equal to each other approximately. Also, $G^{\uparrow\downarrow}$ and $G^{\uparrow\downarrow}$ are small.

## IV. Summary and conclusion

Using tight binding non-equilibrium Green's function method, we have studied the spin-dependent electron transport properties of a silicene ribbon. It has been shown that spin-polarization of transmitted electrons can take place due to the intrinsic SOI and intrinsic Rashba SOI. Then, the effects of an external electric field applied perpendicularly to the ribbon plane have been studied. The optimum values of electron energy at which the spin polarization changes from down spin polarization to up spin polarization, by changing the electric field, have been obtained. Therefore, the spin polarization of transmitted electrons through the silicene ribbon can be controlled and manipulated by changing the strength of electric field. In other words, the silicene ribbon can work as a controllable spin polarizer. In addition, if one uses the incoming electrons with a specific spin polarization, the silicene ribbon can invert the spin of electrons and can works as a spin inverter or a spin NOT gate. Finally, the effect of exchange magnetic field on the spin polarization has been investigated and shown that the silicene ribbon can be used a nearly perfect spin-filter when an appropriate value of exchange field is applied to the silicene ribbon by ferromagnetic substrate.

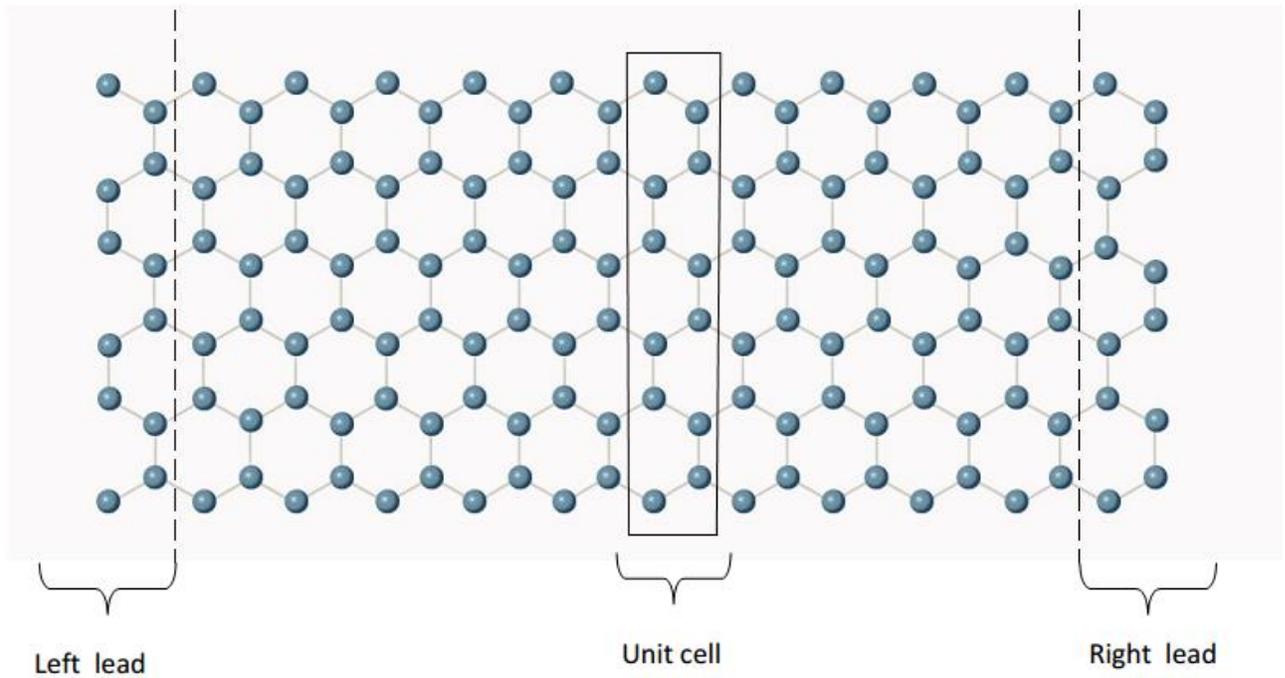

Fig. 1 A buckled silicene ribbon composed by ten unit cells (with total 120 atoms) connected to planar silicene nano-ribbons as left and right leads.

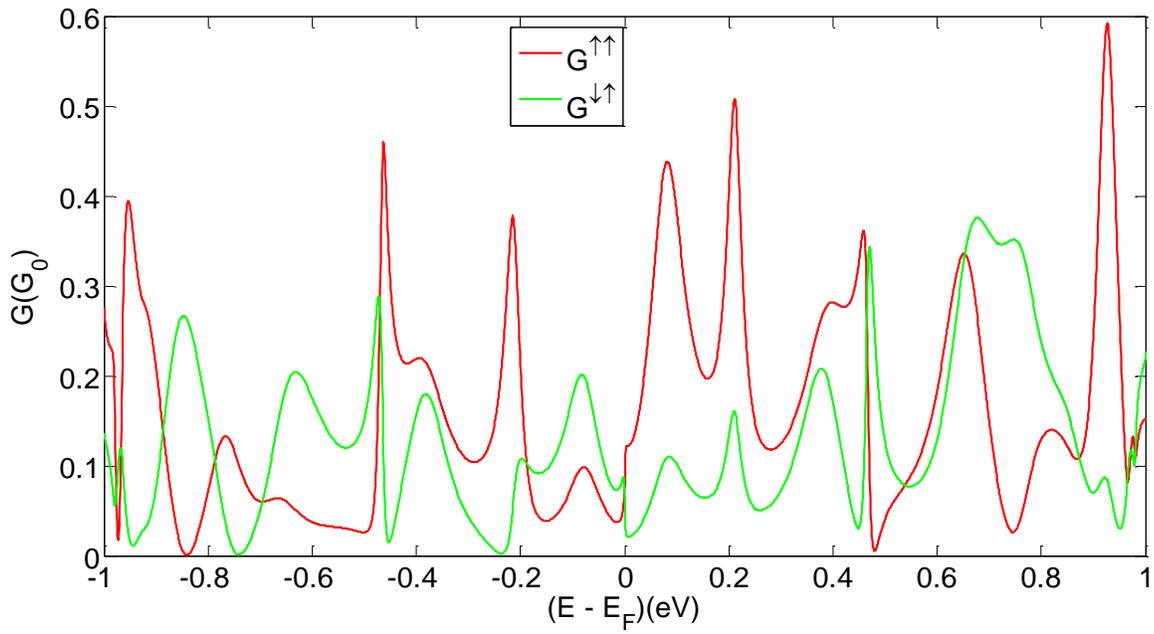

(a)

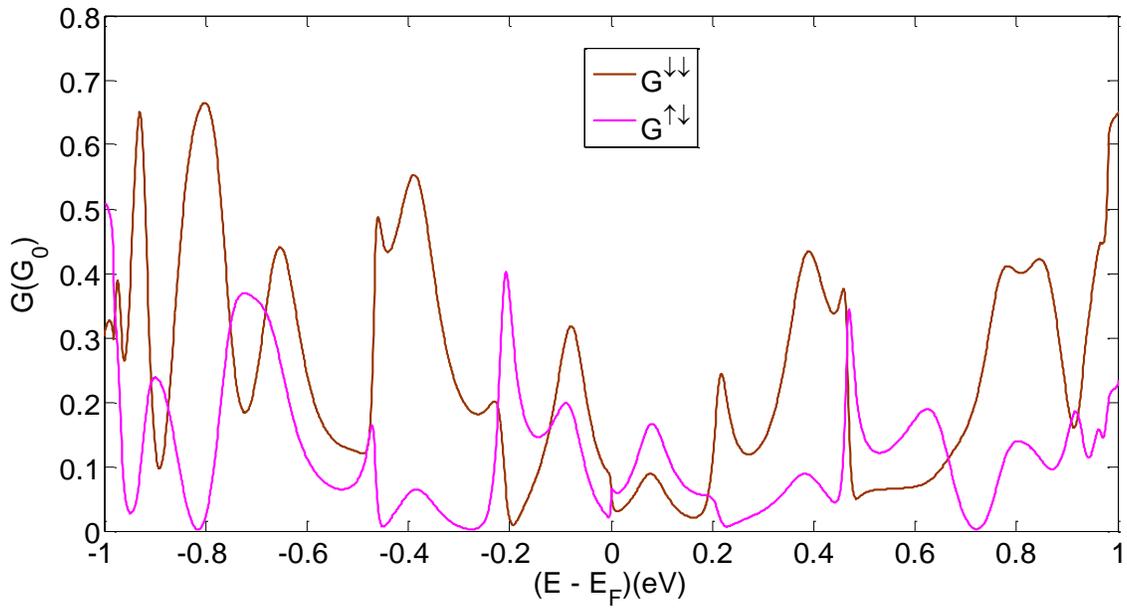

(b)

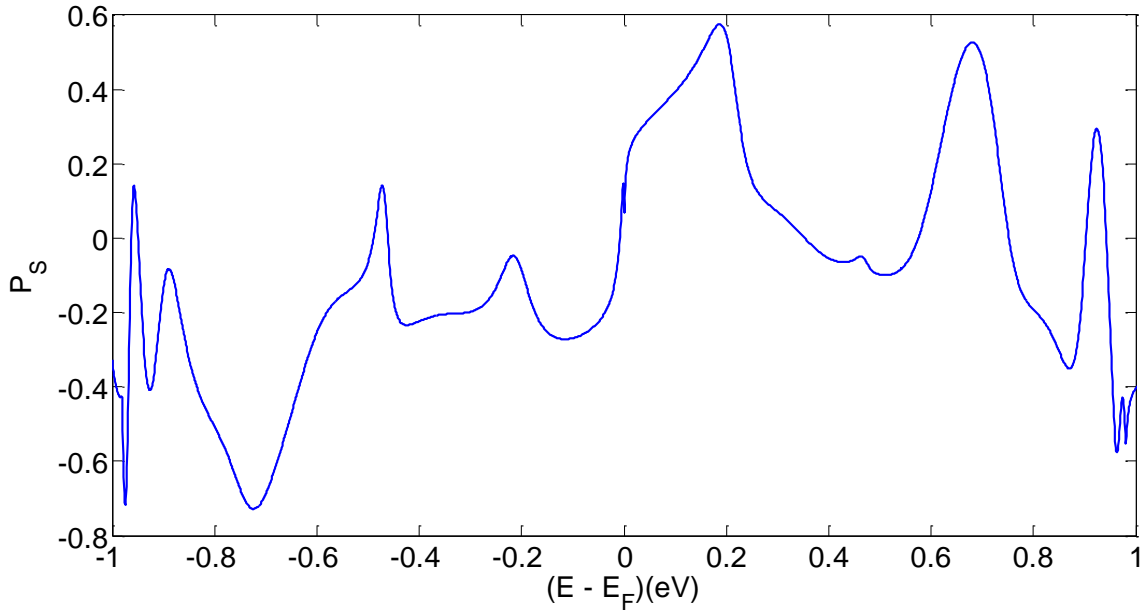

(c)

**Fig. 2** Spin-dependent conductance and spin-polarization of transmitted electrons versus the electron energy. (a) Conductance of transmitted electrons with spin up (i.e., the conductance of spin up without spin flip $G^{\uparrow\uparrow}$ and the conductance of spin down with spin flip to up $G^{\downarrow\uparrow}$). (b) Conductance of transmitted electrons with spin down (i.e., the conductance of spin down without spin flip $G^{\downarrow\downarrow}$ and the conductance of spin up with spin flip to down $G^{\uparrow\downarrow}$). (c) Spin-polarization of transmitted electrons versus the electron energy.

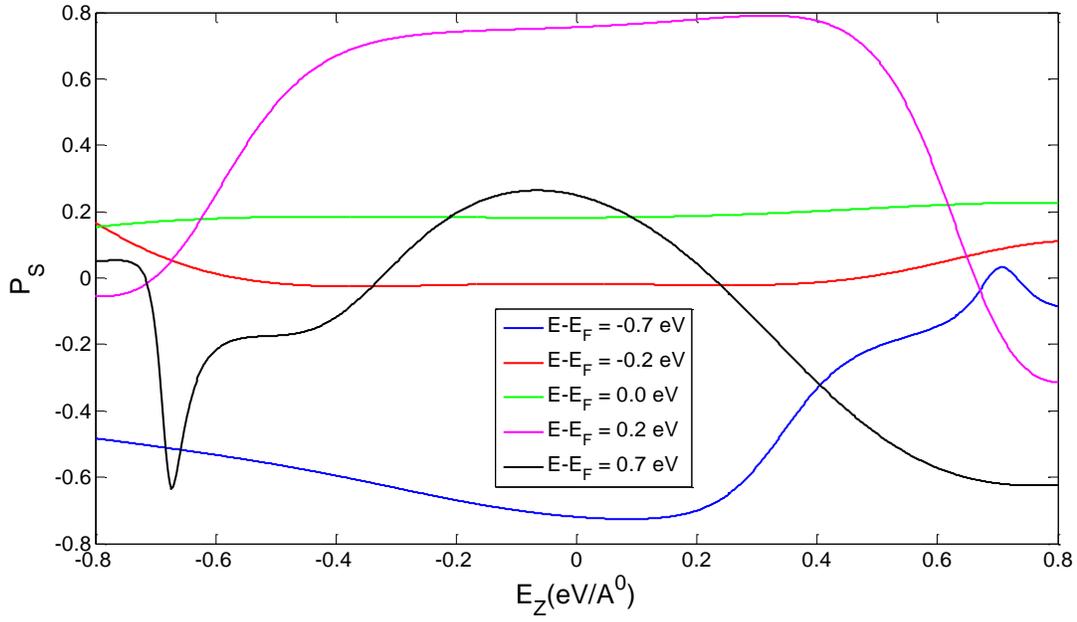

(a)

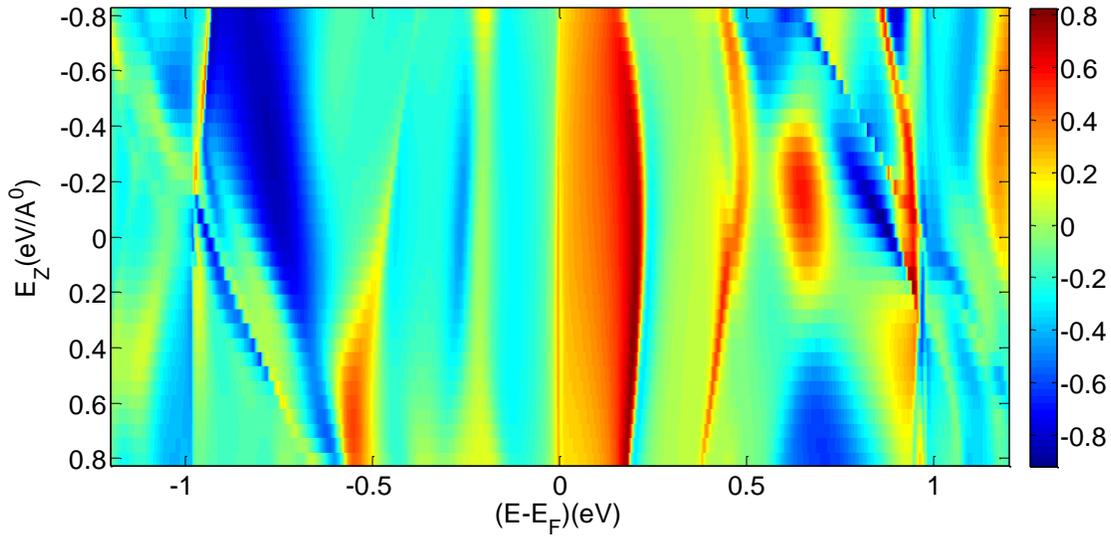

(b)

**Fig. 3** (a) Spin polarization $P_s$ versus the electric field $E_z$ for specific values of electron energy. (b) A three-dimensional graph of spin polarization versus the electron energy and electric field $E_z$.

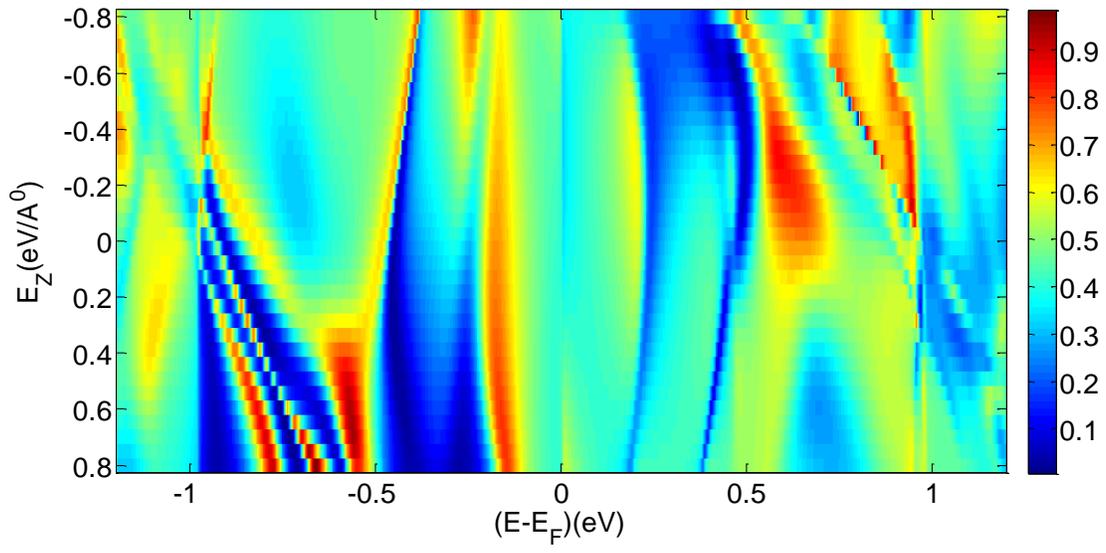

Fig. 4 Three-dimensional graph of spin flipping probability $P_{sf}$ versus the electron energy and electric field $E_z$.

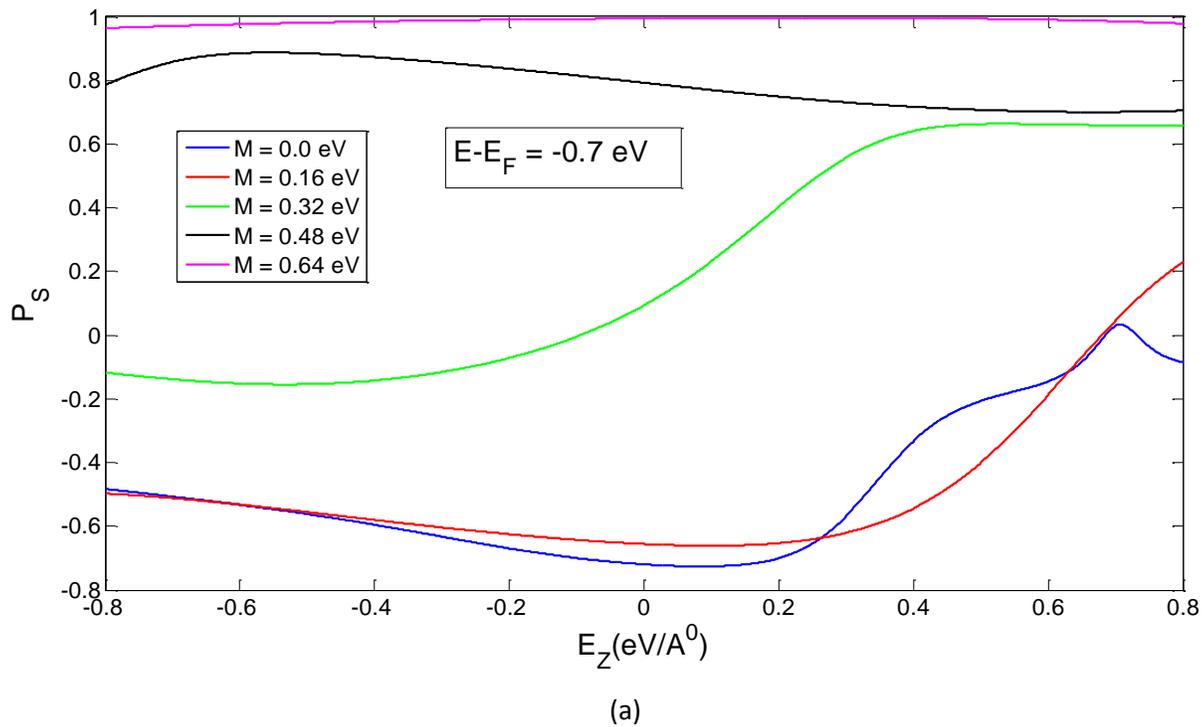

(a)

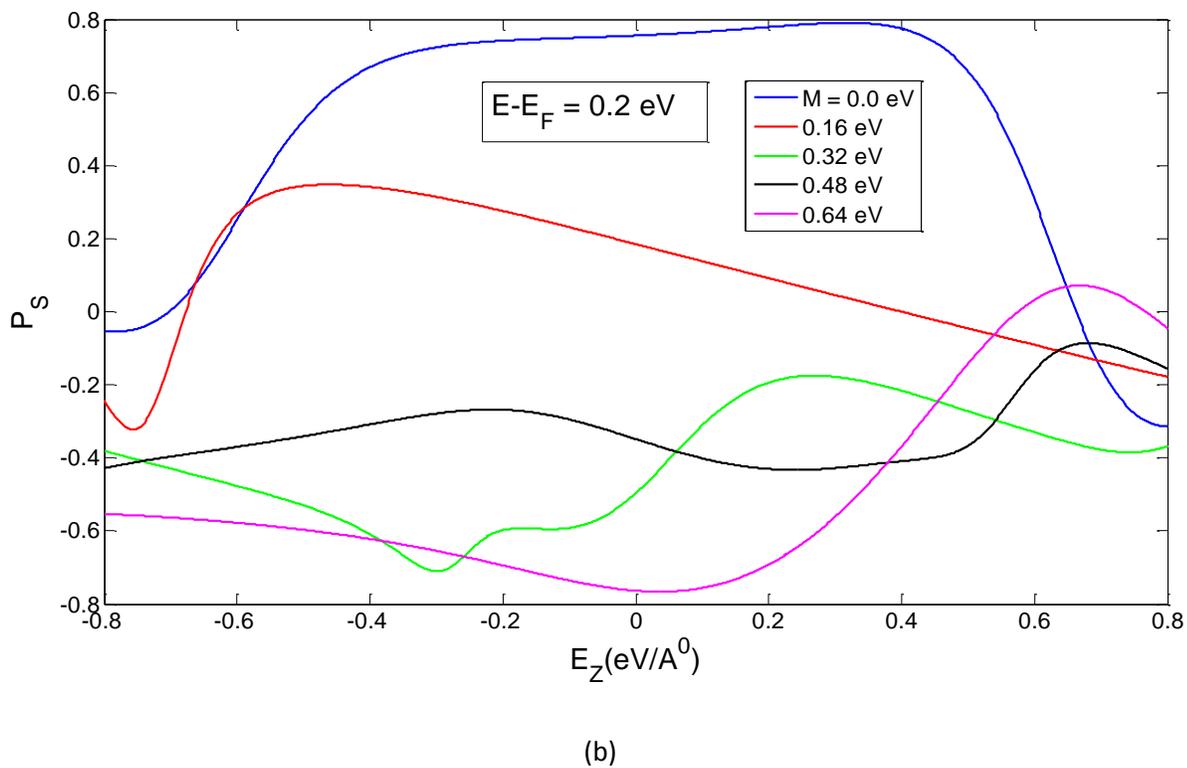

(b)

Fig. 5 Spin polarization versus electric field for (a) $E-E_F$= -0.7 eV and (b) $E-E_F$ = 0.2 eV.

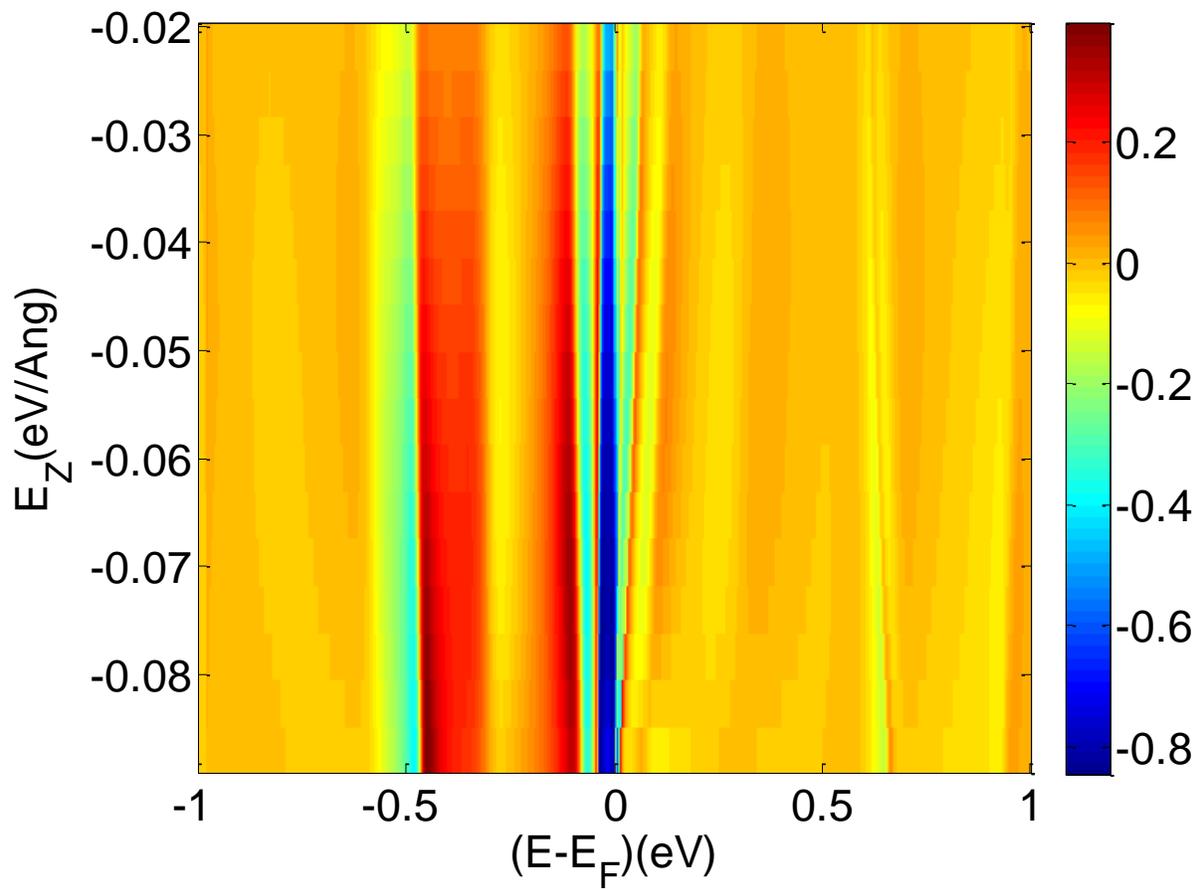

Fig. 6 Three-dimensional graph of spin polarization $P_s$ versus the electron energy and electric field $E_z$. Here, $\lambda_R = 8.66$ meV, and $\lambda_{SO} = 4.2$ meV[32]

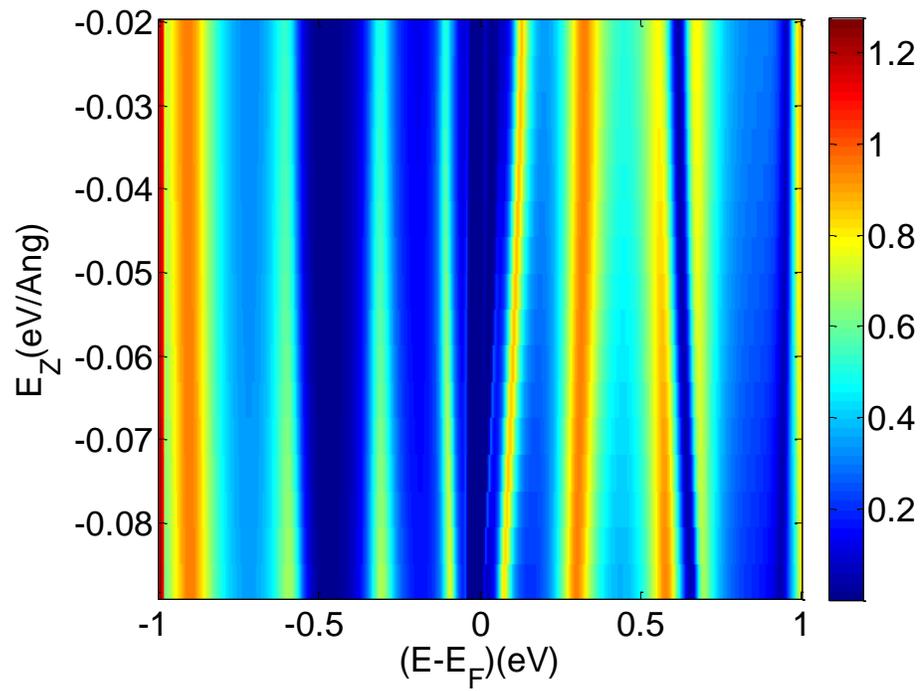

(a)

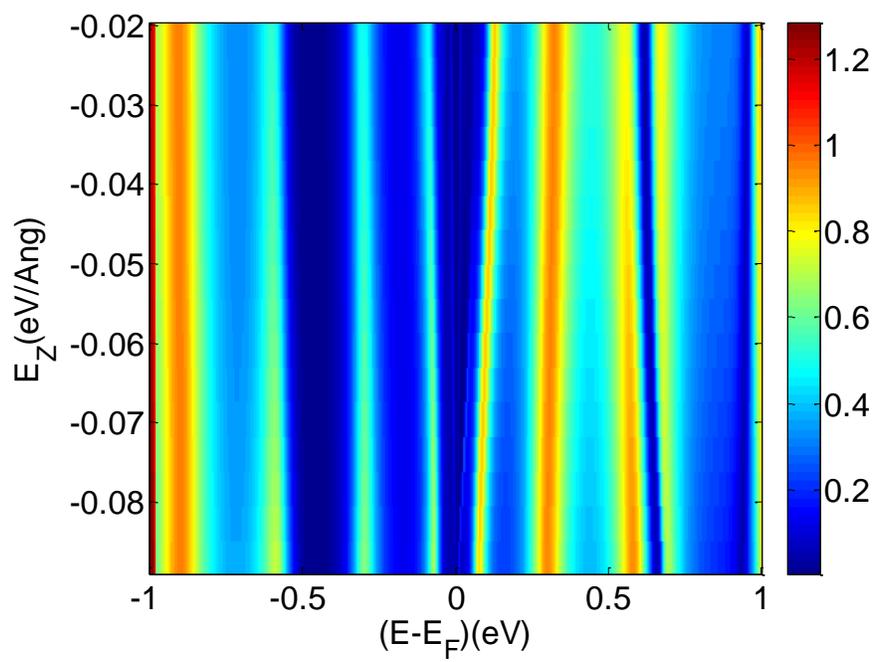

(b)

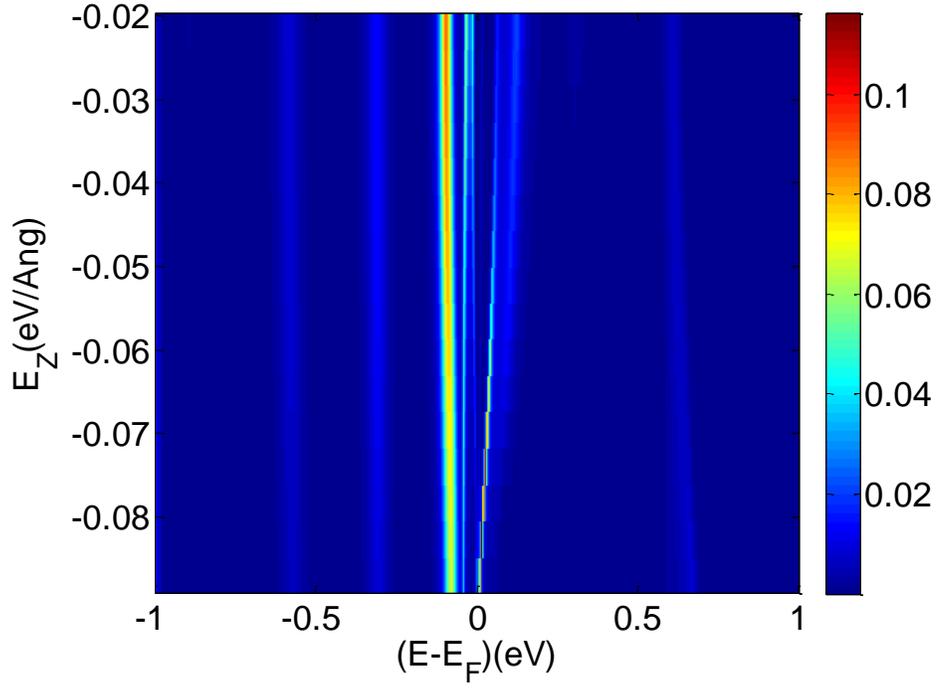

(c)

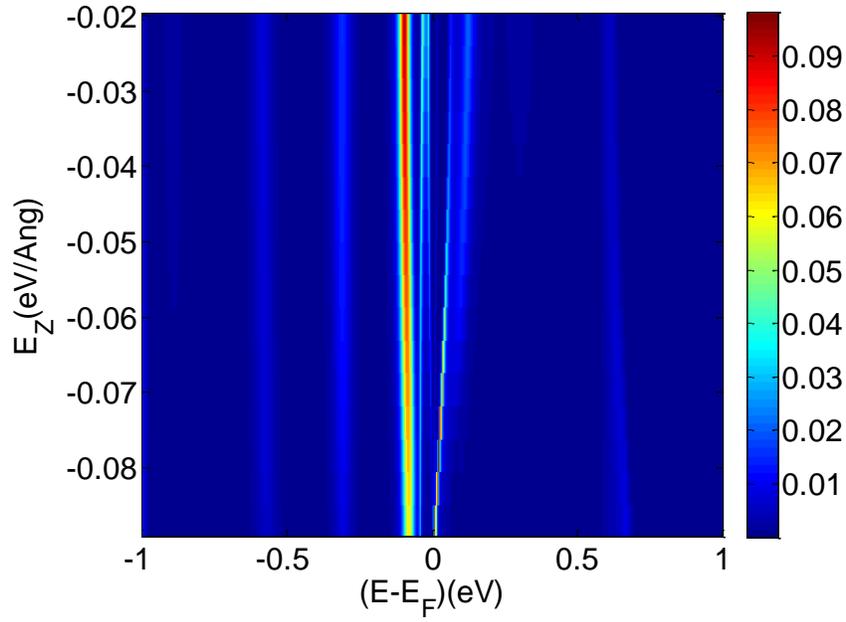

(d)

Fig. 7 Three-dimensional graph of (a) $G^{\uparrow\uparrow}$, (b) $G^{\downarrow\downarrow}$, (c) $G^{\uparrow\downarrow}$ and (d) $G^{\uparrow\downarrow}$ versus the electron energy and electric field $E_z$. Here, $\lambda_R = 8.66$ meV, and $\lambda_{SO} = 4.2$ meV[32]